# Mass shift of σ-Meson in Nuclear Matter

J. R. Morones-Ibarra, Mónica Menchaca Maciel, Ayax Santos-Guevara, and Felipe Robledo Padilla.
Facultad de Ciencias Físico-Matemáticas, Universidad Autónoma de Nuevo León,
Ciudad Universitaria, San Nicolás de los Garza, Nuevo León, 66450, México.
Facultad de Ingeniería y Arquitectura, Universidad Regiomontana,
15 de Mayo 567, Monterrey, N.L., 64000, México.
April 6, 2010

Abstract
The propagation of sigma meson in nuclear matter is studied in the Walecka model, assuming that the sigma couples to a pair of nucleon-antinucleon states and to particle-hole states, including the in medium effect of sigma-omega mixing. We have also considered, by completeness, the coupling of sigma to two virtual pions. We have found that the sigma meson mass decreases respect to its value in vacuum and that the contribution of the sigma omega mixing effect on the mass shift is relatively small.

## 1. INTRODUCTION

The study of matter under extreme conditions of density and temperature, has become a very important issue due to the fact that it prepares to understand the physics for some interesting subjects like, the conditions in the early universe, the physics of processes in stellar evolution and in heavy ion collision. Particularly, the study of properties of mesons in hot and dense matter is important to understand which could be the signature for detecting the Quark-Gluon Plasma (QGP) state in heavy ion collision, and to get information about the signal of the presence of QGP and also to know which symmetries are restored [1].

The study of meson propagation in medium is important to understand properties of nuclear matter [2]. In this work we use the Walecka model to investigate the sigma meson propagation in nuclear mater. The Walecka model is a renormalizable relativistic quantum field theory for nuclear matter where the degrees of freedom are nucleons that interact through the exchange of scalar and vector mesons: the sigma and the omega, respectively. [3,4]. This model has proved to be very successful in explaining several processes in nuclear theory [5].

A lot of work about hadrons in dense nuclear matter has been done in the past few years. One of the interesting phenomenon of nuclear physics that has been recently explained, fitting the data, by the shift down hypothesis of the meson masses in nuclear matter is the anomalous long life time of the C - 14 nucleus [6].

One of the phenomenological reasons to include the σ-meson as a $\bar{q}q$ degree of freedom is the fact that the linear sigma model has been very successful in reproducing the data for many processes in πN and ππ interactions. On the other hand, theoretical predictions and models which have been successful require the existence of the sigma meson [7]. It plays a fundamental role in understanding the symmetry breaking mechanism in QCD an also in the confinement problem [8].

A subject of a long standing controversy has been the existence of the sigma meson with mass about 500MeV. However, recently it was announced that the σ-meson has been detected in ππ→ ππ processes and also in the decay products of heavy quark systems [9]. So, after the experimental announce of the existence of sigma, the controversy is not about its existence but about its structure [9]. The composition of sigma as a $\bar{q}q$-meson or as a ππ resonant particle is still a subject of debate [10, 11, 12,13]. We assume in this paper, that the sigma is a two pion resonance, according to the particle data Group Collaboration [14].

It is still an open question if σ is a fundamental particle, or if it is a dynamically generated particle. Today the discussion about σ has moved from the speculation about its existence, to the structure or the dynamics of sigma [7].

The relevance of sigma meson in hadron and nuclear physics can be shown looking for its contribution to explanation of phenomena like the following [15]: a) The sigma meson is the responsible for the attractive part, in the intermediate range, of the nuclear force, [4, 16]. Without this force we cannot even explain the existence of nuclei, the bound state of nucleons. b) It participate as an intermediate particle in several nuclear processes and in pion-nucleon scattering [16, 17]. c) In meson-meson interaction the sigma occurs as a resonance, showing up as a pole in the T-matrix [16, 18, 19]. In the sigma model, developed by Gell-Mann and Levy [16,20], the sigma meson plays a role similar to the Higgs particle in the Standard Electroweak Theory [16, 20,21], giving mass to the nucleon when the $SU(2)_L \otimes SU(2)_R$ symmetry is spontaneously broken. Additionally it can be proved, from the quark structure, that the sigma is the chiral partner of the pion [22].

According to [9] the identification of sigma meson was realized in the experiment $D^+ \to \pi^- \pi^+ \pi^+$. The experimental results of this three body decays of charm meson can be fitted assuming an intermediate scalar isoscalar state which can be identified with the σ.

In this work we study the scalar sigma meson in nuclear medium when the meson couples to two nucleons through a Yukawa-type coupling. The interaction of the meson with nucleons includes interactions with the Fermi Sea producing nucleon-hole states and interactions with the Dirac Sea producing nucleon-antinucleon states [3, 4]. The sigma-omega mixing is also considered in the calculations.

We have associated the mass of sigma to the peak position of its spectral function. The σ-meson self-energy is calculated in the one loop level and the propagator is computed by summing over ring diagrams, in the so called Random Phase Approximation (RPA) [23]. To carry out the summation we use the Dyson's equation.

## 2. FORMALISM

Starting from the Walecka model we study the possible modifications of the σ-meson mass in the nuclear matter. For this purpose we consider interactions of sigma with particle-hole states, nucleon-antinucleon excitations in the Dirac sea, and the σ-ω mixing effect. [3, 4, 5]. The sigma- omega mixing is a purely density effect which occurs when a particle-hole pair becomes excited in the medium, and, as we can see in Figure 1, the scalar meson decays into a longitudinal omega meson. After this, the inverse process send it back to the original sigma meson. In [24] the authors studied the mixing effect in dilepton production in relativistic heavy ion experiments.

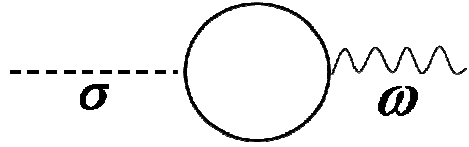

**Figure 1. Sigma - Omega Mixing Diagram**

The study of meson propagation in nuclear medium requires the inclusion of the σ-ω mixing effect. In order to evaluate the mass of σ-meson we need, firstly, to calculate its dressed propagator. One consequence of the mixing effect results in the coupling between the Dyson´s equation for the sigma and omega propagators. The expression for the full sigma-omega propagator can be written as

$$i\Delta(k) = i\Delta_0(k) + i\Delta_0(k)[-i\Sigma(k)]i\Delta(k) \qquad (1)$$

Where $i\Delta_0(k)$ is the non-interacting sigma meson propagator and is written as [3,5]

$$i\Delta_0(k) = \frac{i}{k^2 - \left(m_\sigma^0\right)^2 + i\varepsilon} \qquad (2)$$

and $\Sigma(k)$ is the σ-self-energy. The self-energy $\Sigma(k)$ contains all the information about the interaction between the sigma meson with nucleons and with omega mesons in nuclear matter. In order to determine the self energy $\Sigma(k)$ we must specify the dynamical content of our model.

We start with the Lagrangian density for the Walecka model which is given by [4] as

$$L = \overline{\Psi}[\gamma_\mu(i\partial^\mu - g_v V^\mu) - (M - g_s \Phi)]\Psi + \frac{1}{2}\left(\partial_\mu \Phi \partial^\mu \Phi - (m_\sigma^0)^2\right)\Phi^2 - \frac{1}{4}F_{\mu\nu}F^{\mu\nu} + \frac{1}{2}m_v^2 V_\mu V^\mu + \delta L \qquad (3)$$

where $\Psi$ is the nucleon field with mass $M$, $V^\mu$ is the neutral vector meson field ($\omega$) with mass $m_V$, the tensor field of the vector meson is $F^{\mu\nu} = \partial^\mu V^\nu - \partial^\nu V^\mu$, $\Phi$ is the scalar ($\sigma$) meson field with the bare mass $m_\sigma^0$; $g_s$ and $g_v$ are the coupling constants, and finally $\delta L$ contains renormalization counterterms.

The interaction Lagrangian density which describes the σ-π dynamics, [25] is given by

$$L_{\sigma\pi\pi}^{eff} = \frac{1}{2}g_{\sigma\pi\pi}m_\pi \vec{\pi} \cdot \vec{\pi}\Phi \qquad (4)$$

The influence of the interaction of σ-mesons with nucleons and ω-mesons in nuclear matter is introduced through the modification of the free propagator in the one loop approximation. We will calculate the full propagator in the chain approximation, which consists in an infinite summation of the one loop self-energy diagrams [5]. The diagrammatical representation of the modified propagator is showed in the Figure 2, and the analytical expression is given by $i\Delta(k)$.

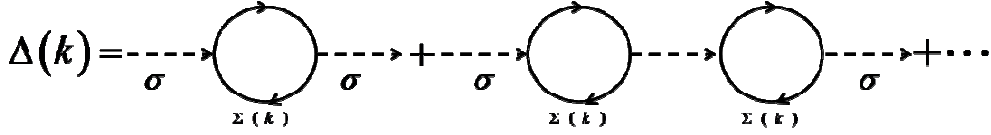

**Figure 2. Sigma meson propagator**

The solution for $\Delta(k)$ in the Dyson's equation (1) is given by

$$\Delta(k) = \frac{1}{\Delta_0^{-1}(k) - \Sigma(k)} = \frac{1}{k^2 - (m_\sigma^0)^2 - \Sigma(k)} \qquad (5)$$

The analytical expression for the self-energy $\Sigma(k)$, is given by [25,26]:

$$\Sigma(k) = \Sigma_{\sigma\pi}(k) + \Sigma_{\sigma N}(k) + \Sigma_\mu^m(k) \qquad (6)$$

Where the calculation of $\Sigma_{\sigma\pi}(k)$ and $\Sigma_{\sigma N}(k)$ are given in [16] and the self-energy of the sigma-omega mixing is

$$i\Sigma_\mu^m(k) = ig_s g_v \int \frac{d^4q}{(2\pi)^4} Tr[G(q)\gamma_\mu G(q+k)] \qquad (7)$$

The scalar-vector mixing effect between σ and ω is a purely medium effect, it occurs when a longitudinal ω meson produces a particle-hole state which decays into a σ, or reciprocally. $G(q) = G_F(q) + G_D(q)$ is the full nucleon propagator [4]:

$$G(q) = (\gamma_\mu q^{*\mu} + M^*)[\frac{1}{q^{*2} - M^{*2} + i\varepsilon} + \frac{i\pi}{E^*(q)}\delta(q_0 - E^*(q))\Theta(k_F - |\vec{q}|))] \qquad (8)$$

where, $q^{*\mu} \equiv (q^0 - g_v V^0, \vec{q})$, $E^*(q) \equiv \sqrt{\vec{q}^2 + M^{*2}}$, $k_F$ is the Fermi momentum, and M* is the nucleon effective mass.

We take as the definition of the mass of a particle the magnitude of the four momentum K for which the spectral function gets its maximum, a definition that is used extensively in the literature, [27,28,29] and it is well established. This is the definition that we will use in this work for calculating the meson mass in nuclear matter.

## 3. CALCULATION OF THE SIGMA-OMEGA MIXING TERM

Integrating Eq.(8) respect to $q^0$ we obtain

$$\Sigma_\mu^m(k) = 8\pi g_s g_v \int \frac{d^3q}{(2\pi)^4} \frac{[4k^\mu q^2 + 4k \cdot q q^\nu]\Theta(k_F - |\vec{q}|)}{[(k^2)^2 - 4(q \cdot k)^2]E^*(q)} \quad (9)$$

On the other hand, as the barionic current is conserved it follows that

$$k^\mu \Sigma_\mu^m(k) = 0 \quad (10)$$

Taking the condition as for low energies [3] and taking the $\vec{k}$ in the positive direction of the z axis, we find

$$\Sigma_\mu^{Rm}(k) = \frac{g_s g_v M^*}{\pi^2}[2k_F - C_0\varepsilon_F \ln\left|\frac{C_0\varepsilon_F + k_F}{C_0\varepsilon_F - k_F}\right|] \quad (11)$$

where $C_0 = \frac{k_0}{|\vec{k}|}$ and $\varepsilon_F \equiv \sqrt{k_F^2 + M^{*2}}$.

The propagator takes the form

$$\Delta(k^2) = \frac{1}{k^2 - m_\sigma^2 - \text{Re}\Sigma^R(k^2) - i\text{Im}\Sigma^R(k^2)} \quad (12)$$

with

$$\text{Re}\Sigma^R(k) = \text{Re}[\Sigma_{\sigma\pi}^R(k) + \Sigma_{\sigma N}^R(k) + \Sigma_\mu^{Rm}(k)] \quad (13)$$

$$\text{Im}\Sigma(k) = \text{Im}\Sigma_{\sigma\pi}(k) \quad (14)$$

The parameters in Eq.(12) are the reported $m_\sigma$ value for the σ-meson mass, which we take in the range $450\text{MeV} \leq m_\sigma \leq 550 \text{ MeV}$, the bare σ ππ coupling constant $g_{\sigma\pi\pi} = 12.8$, the bare σ N coupling constant as $g_s^2 = 54.289$ and the bare ωN coupling constant as $g_v^2 = 102.770$ [5].

The spectral function of the sigma meson at two times the normal nuclear matter density has been plotted in Figure 3. The nucleon mass was fixed at their physical value M = 939MeV. The effective nucleon mass M* is the appropriate value at nuclear-matter saturation density. This relation was taken as M* / M = 0.730, and the bare σ N coupling constant $g_s^2 = 54.289$ [5].

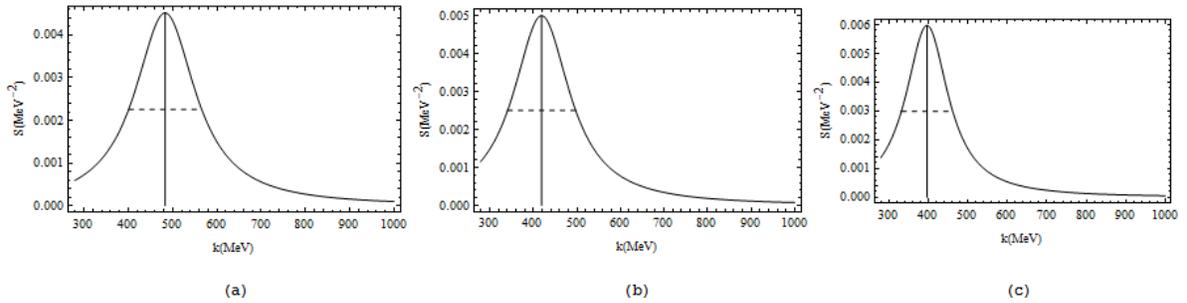

(a)      (b)      (c)

**Figure 3. Spectral function of the sigma meson at two times the normal nuclear matter: a) in vacuum, b) without mixing effect and c) with mixing effect**

As it is showed in the graphs, the value of the mass of sigma in medium suffer a down shift respect to its value in vacuum. It is worth to say that there is not a unique result concerning the shift mass of mesons in medium. In [2] they obtain an increasing in the effective mass of sigma. On the other hand, others authors, by using different models as Nambu-Jona-Lasinio or QCD sum rules, have obtaining a reduction in medium for the sigma meson effective mass.

## 4. CONCLUSION

In the study of the propagation of the scalar sigma meson in nuclear matter we have found that the effective mass of the meson decreases respect to its value in vacuum. Assuming that the σ couples to a particle-hole pair in nuclear matter, that it interacts to nucleon-antinucleon states in the Dirac sea, and also by including, by completeness, the coupling of sigma to two pions, we obtained 13% of reduction in the value of the sigma meson mass without considering the scalar-vector mixing and 18% of reduction considering the mixing effect. The mixing effect between σ and ω occurs only in medium, when a longitudinal ω meson produces a particle-hole state which decays into a σ , or reciprocally. When this sigma-omega mixing is included, we found that the effect is small on the sigma effective mass . The value of the effective mass of sigma when the σ -ω mixing effect is included is reduced a little bit more, a difference of the 5% at two times the normal nuclear matter density. These results are in agreement with those obtained by using other models, like the Nambu Jona-Lasinio [30,31] and the chiral perturbation theory [18, 19], which predict a decreasing of the mass of the sigma meson in medium. This result can be interpreted as a partial restoration of chiral symmetry [32,33]. On the other hand, taking the width at one half of the maximum value of the spectral function, we obtained for it a of reduction of 23% respect to the width in the vaccum.